\documentclass[proceedings]{JHEP} 

\usepackage{epsfig}                     

\newbox\mybox
\newcommand\fverb{\setbox\mybox=\hbox\bgroup\verb}
\newcommand\fverbdo{\egroup\medskip\noindent\fbox{\unhbox\mybox}\ }
\newcommand\fverbit{\egroup\item[\fbox{\unhbox\mybox}]}

\font\beeg=cmr17 scaled 1600            
\newcommand\init[1]{\setbox\mybox=\hbox{{\beeg #1}~}%
                   \noindent\global\hangindent=\wd\mybox\global\hangafter-2%
                   \sc\smash{\llap {\lower 13.2pt \box\mybox}}}

\title{Conceptual aspects of QCD factorization 
in hadronic $B$ decays\footnote{
Talk presented at the UK Phenomenology Workshop on Heavy Flavour 
and CP Violation, 17 - 22 September 2000, St John's College, Durham, 
proceedings to appear in J. Phys. G. 
Also presented at: 
5th International Symposium on Radiative Corrections (RADCOR2000), 
Carmel, California, September 11 - 15, 2000;
4th Workshop on Continuous Advances in QCD, Minneapolis, U.S.A., 
12-14 May 2000;
Vth International Workshop on Heavy Quark Physics, Dubna, Russia, 
6-8 April 2000. 
The phenomenological parts of some of the talks have been omitted in this 
write-up. See Ref.~\cite{BBNS00b} for those parts.}}

\author{M.~Beneke\\
        Institut f\"ur Theoretische Physik E, RWTH Aachen,
Sommerfeldstr.~28, 52074 Aachen, Germany\\
        Email: \email{mbeneke@physik.rwth-aachen.de}}

\conference{PITHA 00/23, hep-ph/0009328, September 2000}

\abstract{I review the meaning of ``QCD factorization'' 
in hadronic two-body $B$ decays and then discuss recent results 
of theoretical (rather than phenomenological) nature: the proof 
of factorization at two loops; the identification of ``chirally 
enhanced'' power corrections; and the role of annihilation 
contributions.}

\begin{document} 

\section{Introduction}

Hadronic, two-body $B$ decays are highly interesting observables 
for flavour physics, since they depend on CKM matrix elements, 
including the CP-violating phase of the CKM matrix, and potential 
other flavour-changing interactions. They also present a formidable 
challenge for theory, since they involve three fundamental scales, the weak 
interaction scale $M_W$, the $b$-quark mass $m_b$, and the QCD  
scale $\Lambda_{\rm QCD}$. From the point of view of fundamental 
physics, the sensitivity to the weak interaction scale, and potential 
new phenomena at this scale, is probably most interesting, but since 
this physics is weakly coupled, it is straightforwardly computable, 
given a particular model of flavour violation. Most theoretical work 
therefore concerns strong-interaction corrections. The 
strong-interaction effects which involve virtualities above the 
scale $m_b$ are well understood. They renormalize the coefficients of 
local operators ${\cal O}_i$ in the weak effective
Hamiltonian. Assuming the Standard Model of flavour violation,  
the amplitude for the decay $B\to M_1 M_2$ is given by
\begin{eqnarray}
\label{effham}
{\cal A}(B\to M_1 M_2) &=& 
\nonumber\\
&&\hspace*{-2.5cm}
\frac{G_F}{\sqrt{2}} \sum_i 
\lambda_i\,C_i(\mu)\,
\langle M_1 M_2 |{\cal O}_i|B\rangle(\mu),
\end{eqnarray}
where $G_F$ is the Fermi constant. Each term in the sum is 
the product of a CKM factor $\lambda_i$, 
a coefficient function $C_i(\mu)$, which incorporates strong-interaction 
effects above the scale $\mu\sim m_b$, and a matrix element of an 
operator ${\cal O}_i$. In extensions of the Standard Model, there may 
be further operators and different flavour-violating couplings, but 
the strong-interaction effects below the scale $\mu$ are still 
encoded by matrix elements of local operators. 

The theoretical problem is therefore to compute these matrix 
elements. Since they depend on $m_b$ and $\Lambda_{\rm QCD}$, one 
should take advantage of the fact that $m_b\gg \Lambda_{\rm QCD}$ 
and compute the short-distance part of the matrix element. The 
remainder then depends only on $\Lambda_{\rm QCD}$, and -- to 
leading order in $\Lambda_{\rm QCD}/m_b$ -- turns out 
to be much simpler than the original matrix element. In this talk 
I summarize some conceptual aspects of our recent 
work~\cite{BBNS99,BBNS00a} on this problem. 
The discussion of the phenomenology of some 
particular decay modes is omitted here, but can also be found 
in Refs.~\cite{BBNS99,BBNS00a,BBNS00b}. 

\section{QCD Factorization}

\subsection{The Physical Picture}

Factorization is a property of the heavy-quark limit, in which we assume 
that the $b$ quark mass is parametrically large. The $b$ quark is then 
decaying into a set of very energetic partons. How these partons 
and what is left of the $B$ meson hadronize into two mesons depends 
on the identity of these mesons.

The simplest case is $\bar{B}_d\to D^+\pi^-$, when the $D$ meson is 
also taken to be parametrically heavy. The spectator quark and other 
light degrees of freedom in the $B$ meson have to rearrange themselves 
only slightly to form a $D$ meson together with the charm quark 
created in the weak interaction. The other two light quarks are 
very energetic and for them to form a pion they must be highly 
collinear and in a colour-singlet configuration. Soft interactions 
decouple from such a configuration and this allows it to leave 
the decay region without interfering with the $D$ meson formation. 
The probability of such a special configuration to form a pion 
is described by the leading-twist pion light-cone distribution 
amplitude (LCDA) $\Phi_\pi(u)$. The $B\to D$ transition is parameterized 
by a standard set of form factors. I have repeated essentially 
the argument of Ref.~\cite{Bj89} in favour of the conventional 
factorization picture, but it is important that this can be converted 
into a quantitative scheme to compute higher order corrections. 
For example, if the light quark-anti-quark pair is initially formed 
in a colour-octet state, we can still show that soft gluons 
decouple, if this pair is to end up as a pion. This implies that 
the pair must interact with a hard gluon, and hence this provides a 
calculable strong-interaction correction to the basic mechanism 
discussed above. (A correction of this type was computed already in 
Ref.~\cite{PW91}, but it seems to me that the generality and 
importance of the result went unnoticed.) An important element in 
demonstrating the suppression of soft interactions (except for 
those parameterized by the $B\to D$ form factor) is the assumption 
that the pion LCDA vanishes linearly as the longitudinal momentum 
fraction approaches the endpoints $u=0,1$. This assumption can be 
justified by the fact that it is satisfied 
by the asymptotic distribution amplitude $\Phi_\pi(u)= 6 u (1-u)$, 
which is the appropriate one in the heavy-quark limit.
As a consequence
\begin{equation}
\label{endpoints}
\int_0^{\Lambda_{\rm QCD}/m_b} \!\!\!\!\!\!\!\!du \, u^n\,\Phi_\pi(u) 
\sim \left(\frac{\Lambda_{\rm QCD}}{m_b}\right)^{n+2}
\quad \!\!\!\!\!\!\!(n>-2).
\end{equation}
This guarantees the suppression of soft endpoint contributions. 

Note that the above discussion relies crucially on the spectator 
quark going to the heavy meson in the final state. If, as in the 
case of a $D^0 \pi^0$ final state, the spectator quark must be 
picked up by the light meson, the amplitude is suppressed by the 
$B\to \pi$ form factor. But since the $D$ meson's size is of 
order $1/\Lambda_{\rm QCD}$, the $D^0$ formation and $B\to \pi$ 
transition cannot be assumed to not interfere and factorization 
is violated. 

The case of two light final state mesons is the most interesting 
one. The dominant decay process is indeed the same as for the 
case of the $D^+ \pi^-$ final state, but this implies that the 
light meson that picks up the spectator quark is formed in a very 
asymmetric configuration in which the spectator quark carries a tiny 
fraction $\Lambda_{\rm QCD}/m_b$ of the total energy. Such a 
configuration is suppressed, see (\ref{endpoints}), and this
suppression is equivalent to the well-known 
$(\Lambda_{\rm QCD}/m_b)^{3/2}$-suppression~\cite{CZ90} of 
heavy-to-light form factors at large recoil. Owing to this
suppression there exists a competing process, in which a hard 
gluon is exchanged with the spectator quark, propelling it to 
large energy, thus avoiding the penalty factor of 
(\ref{endpoints}). If the hard gluon connects to the 
quark-antiquark pair emanating from the weak decay vertex to form 
the other light meson, this gives rise to another contribution 
to the factorization formula. (If the gluon connects to the $b$ quark 
or the quark that forms the light meson together with the spectator 
quark, we can consider this as a hard-scattering contribution to the 
heavy-to-light form factor.) This further contribution, called 
``hard-spectator interaction'', can be computed with standard 
methods for light-cone-dominated reactions~\cite{LB80,EfRa80}. 

There also exist ``annihilation'' contributions, defined as those 
diagrams, in which the spectator fermion line connects to the weak 
decay vertex. These contributions are 
suppressed by the factor
\begin{equation}
\label{lamb}
\frac{\int d\xi\,\Phi_B(\xi)}{\int d\xi\,\Phi_B(\xi)/\xi} 
\equiv \frac{\lambda_B}{M_B} \sim\frac{\Lambda_{\rm QCD}}{m_b},
\end{equation}
where $\Phi_B(\xi)$ is the $B$ meson LCDA and 
$\xi\sim \Lambda_{\rm QCD}/m_b$ the light-cone momentum fraction 
of the spectator quark. Hence annihilation contributions can be 
neglected in the heavy-quark limit (but see the later 
discussion).

\subsection{The Factorization Formula}

\begin{figure}[t]
   \vspace{-1.5cm}
   \epsfysize=13.5cm
   \epsfxsize=9cm
   \centerline{\epsffile{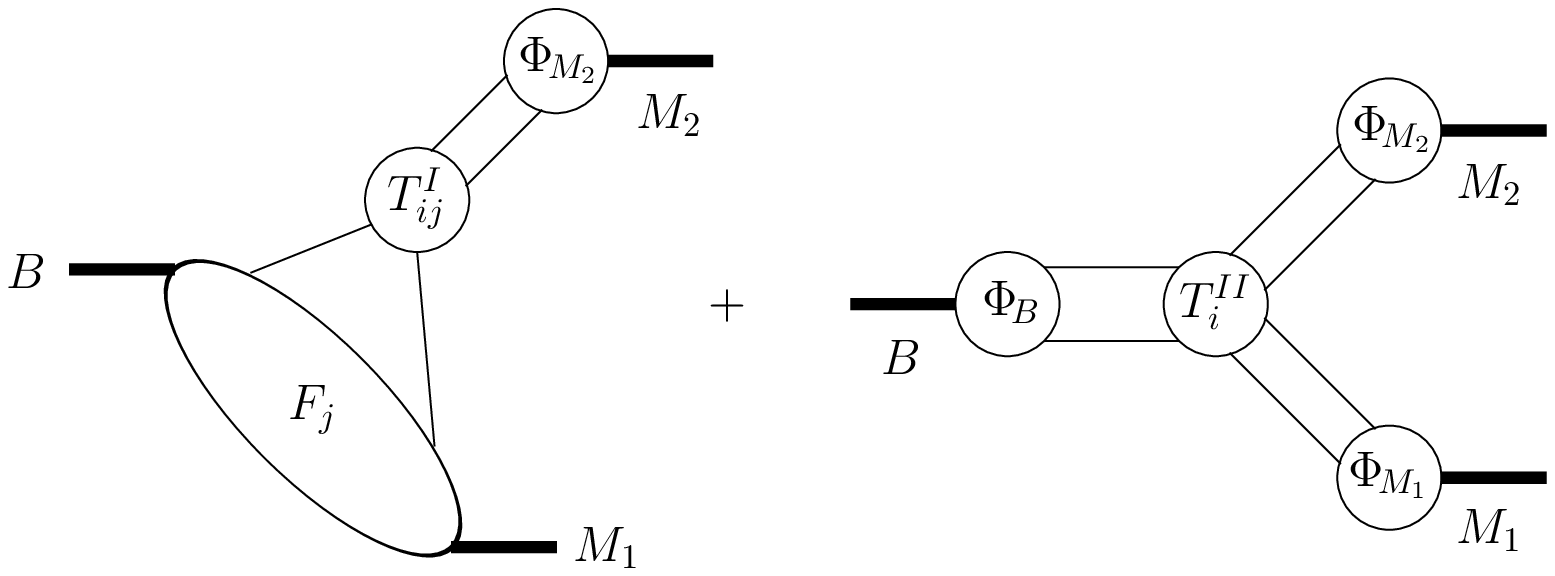}}
   \vspace*{-9cm}
\caption[dummy]{\label{fig1}\small Graphical representation of 
the factorization formula. Only one of the two form-factor terms in 
(\ref{fff}) is shown for simplicity.}
\end{figure}

We consider weak decays $B\to M_1 M_2$ in the heavy-quark limit. The 
formal expression of the previous discussion is given by the following
result for the matrix element of an operator ${\cal O}_i$ in the weak 
effective Hamiltonian, valid up to corrections of order 
$\Lambda_{\rm QCD}/m_b$:
\begin{eqnarray}
\label{fff}
\langle M_1 M_2|{\cal O}_i|\bar{B}\rangle &=& 
\nonumber\\
&&\hspace*{-2.3cm}
\sum_j F_j^{B\to M_1}(m_2^2)\,\int_0^1 du\,T_{ij}^I(u)\,\Phi_{M_2}(u) 
\nonumber\\
&&\hspace*{-2.3cm}
+\,\,(M_1\leftrightarrow M_2)\nonumber\\
&&\hspace*{-2.3cm}
+\int_0^1 \!d\xi du dv \,T_i^{II}(\xi,u,v)\,
\Phi_B(\xi)\,\Phi_{M_1}(v)\,\Phi_{M_2}(u)  \nonumber\\
&&\hspace*{-1.5cm}\mbox{if $M_1$ and $M_2$ are both light,}\\
\langle M_1 M_2|{\cal O}_i|\bar{B}\rangle &=& 
\nonumber\\
&&\hspace*{-2.3cm}
\sum_j F_j^{B\to M_1}(m_2^2)\,\int_0^1 du\,T_{ij}^I(u)\,\Phi_{M_2}(u) 
\label{fff2}
\nonumber\\
&&\hspace*{-1.5cm}\mbox{if $M_1$ is heavy and $M_2$ is light.} 
\end{eqnarray} 
Here $F_j^{B\to M_{1,2}}(m_{2,1}^2)$ denotes a $B\to M_{1,2}$ form factor, 
and $\Phi_X(u)$ is the light-cone distribution amplitude for the 
quark-antiquark Fock state of meson $X$. 
$T_{ij}^I(u)$ and $T_i^{II}(\xi,u,v)$ are perturbatively calculable 
hard-\-scattering functions;  
$m_{1,2}$ denote the light meson masses.  
Eq.~(\ref{fff}) is represented graphically in 
Fig.~\ref{fig1}. (The fourth line of (\ref{fff}) is 
somewhat simplified and may require including an integration over 
transverse momentum in the $B$ meson starting from order 
$\alpha_s^2$.)

Eq.~(\ref{fff}) applies to decays into two 
light me\-sons, for which the spectator quark in the $B$ meson 
can go to either of the final-state mesons. An example is the decay 
$B^-\to \pi^0 K^-$. If the spectator quark can go only to one of the 
final-state mesons, as for example in $\bar B_d\to \pi^+ K^-$, 
we call this meson $M_1$ and the second form-factor term 
on the right-hand side of (\ref{fff}) is absent. 
The factorization formula simplifies when the spectator quark goes 
to a heavy meson [see (\ref{fff2})], 
such as in $\bar{B}_d\to D^+ \pi^-$. In this case 
the hard interactions with the spectator quark can be dropped because
they are power suppressed in the heavy-quark limit. 
In the opposite situation that 
the spectator quark goes to a light meson but the other meson is heavy, 
factorization does not hold as discussed above. 

As an example, consider the matrix element $\langle\pi^+\pi^-|
(\bar{u}b)_{V-A}(\bar{d}u)_{V-A}|\bar{B}\rangle$. In leading order 
the conventional factorization result $i f_\pi F^{B\to \pi}_+(0)
M_B^2$ is obtained. It is convenient to introduce ``factorized operators'' 
$j_1\otimes j_2$, whose matrix elements are {\em defined} by 
$\langle \pi\pi |j_1\otimes j_2|B\rangle\equiv
\langle\pi|j_1|B\rangle\,\langle \pi|j_2|0\rangle$. The benefit of 
this notation is that the result of the factorization formula can be 
expressed in terms of these factorized operators, and this gives 
a compact representation of the result for {\em all} $\pi\pi$ final 
states. We then find, including the order-$\alpha_s$ corrections,  
\begin{eqnarray}
(\bar{u}b)_{V-A}(\bar{d}u)_{V-A} &=& 
(\bar{u}b)_{V-A}\otimes (\bar{d}u)_{V-A}
\nonumber\\[0.1cm]
&&\hspace*{-3cm}+ \frac{1}{3}\,
(\bar{d}b)_{V-A}\otimes (\bar{u}u)_{V-A}
\nonumber\\[0.1cm]
&&\hspace*{-3cm}+\, \frac{\alpha_s}{9\pi} \bigg[
 \left\{V+H\right\}\,(\bar{d}b)_{V-A}\otimes (\bar{u}u)_{V-A}
\nonumber\\[0.1cm]
&&\hspace*{-3cm}+P \sum_{q=u,d}(\bar{q}b)_{V-A}\otimes (\bar{d}q)_{V-A}\bigg],
\end{eqnarray}
where 
\begin{eqnarray}
V &\equiv& \int_0^1 \!d u \,\Phi_\pi(u) 
\bigg[6\ln\frac{m_b^2}{\mu^2}-18
\nonumber\\
&&\hspace*{0cm}+\,\frac{3 (1-2 u)}{1-u}\ln u-3 \pi
i\bigg],
\\[0.2cm]
P&\equiv& \int_0^1 \!d u \,\Phi_\pi(u) 
\bigg[ \frac{2}{3}\ln\frac{m_b^2}{\mu^2}
    + \frac{2}{3} \\
&&\hspace*{0cm}+\,4\int_0^1\!dz\,z(1-z)
    \ln[-z(1-z)u-i\epsilon]\bigg]
\nonumber\end{eqnarray}
correspond to the first line of (\ref{fff}), which includes vertex 
($V$) and penguin ($P$) contractions and 
\begin{eqnarray}
\label{hard}
H &\equiv& \frac{4\pi^2}{3}\,\frac{f_B f_\pi}{M_B \lambda_B F^{B\to \pi}_+(0)}
    \int_0^1 \frac{du}{u}\,\Phi_\pi(u)
\nonumber\\
\hspace*{-2cm}
&&\times\,\int_0^1 \frac{dv}{v} \left[ 
    \Phi_\pi(v) + \frac{2\mu_\pi}{m_b}\,\Phi_p(v)\right], 
\end{eqnarray}
accounts for the hard spectator scattering, the second line of 
(\ref{fff}). (I have included a certain power 
correction, proportional to a twist-3 LCDA $\Phi_p(v)$, 
which will be discussed in some detail later.) After evaluating the 
matrix elements of all operators in the weak effective hamiltonian, 
we collect the coefficients of the factorized operators into a set 
of numbers, conventionally denoted $a_i$ and express the decay
amplitude in terms of these $a_i$. In the past, these $a_i$ have 
been thought to be uncalculable and have often been assumed to be 
universal. We now see that they are calculable, but non-universal to
the extent that they depend on the identity of the meson.

In the past there have been several attempts at describing 
exclusive $B$ decays in terms of hard spectator scattering 
alone~\cite{SHB90,BuDo91,SiWy91,CaMi93,ASY94,Ward1995,DaJK95,Li:1996iu} 
and recently these have been revived in 
Refs.~\cite{Keum:2000wi,Lu:2000em}. The underlying assumption here is that the 
$B\to \pi$ form factors are themselves computable in this approach.  
The literature quoted reaches different conclusions on the 
validity of this assumption. The principal difficulty lies in 
the appearance of an endpoint divergence in the integration 
over the pion LCDA. This must either be interpreted as an indication 
of a soft contribution to the form factor, in which case the starting 
assumption is invalid; or the divergence must be argued 
away (either by subtracting it into the $B$ meson wave function~\cite{ASY94}
or by suppressing it with a Sudakov form factor~\cite{Li:1996iu}), 
in which case the typical (but not unequivocal) 
result is an unacceptably small 
form factor at $q^2=0$. In comparison 
the advantage of the approach presented here 
is that it allows us to demonstrate a useful factorization result 
independent of the validity of the assumption that the form factor 
is dominated by hard scattering. (I may mention that form factors 
themselves obey a formula similar to (\ref{fff}) and this 
may provide a useful framework to investigate the role of 
hard scattering for form factors~\cite{BF00}.)

\subsection{Implications of Factorization}

The significance and usefulness of the factorization formula 
stems from the fact that the non-perturbative quantities which appear 
on the right-hand side of (\ref{fff}) 
are much simpler than the original non-leptonic 
matrix element on the left-hand side. This is because they either 
reflect universal properties of a single meson state (light-cone 
distribution amplitudes) or refer only to a $B\to\mbox{meson}$ transition 
matrix element of a local current (form factors). Factorization 
is a consequence of the fact that only {\em hard\,} interactions 
between the $(B M_1)$ system and $M_2$ survive in the heavy 
quark limit. As a result we can say that

\begin{itemize}
\item[-] conventional factorization gives the correct limit, when 
$\alpha_s$ and $\Lambda_{\rm QCD}/m_b$ corrections are neglected, 
provided the spectator quark does not go to a heavy meson;
\item[-] radiative corrections to conventional factorization can be 
computed systematically, the result is, in general, non-universal, 
i.e. there is no reason to suppose that the parameters $a_i$ 
should be the same, say for $D\pi$ and $\pi\pi$ final states;
\item[-] the problem of scheme-dependence in the conventional 
factorization ansatz is solved in the same way as in any other 
next-to-leading order computation with the weak effective 
hamiltonian;
\item[-] all strong 
interaction phases are generated perturbatively in the heavy 
quark limit, as form factors have no imaginary parts;
\item[-] many observables of interest for CP violation 
become accessible in this way, the current limiting factors being 
our poor knowledge of $\lambda_B$ and that of power corrections.
\end{itemize}

\section{Discussion}

In this section I discuss some aspects that concern factorization  
beyond the one-loop correction to conventional factorization: the validity 
of factorization in higher orders of perturbation theory 
[Sect.~\ref{facthigh}], the issue of final state rescattering 
[Sect.~\ref{rescatt}] and various sources of power corrections 
[Sects.~\ref{higherfock}-\ref{annihilation}].

\subsection{Factorization in higher orders}
\label{facthigh}

A proof of the factorization formula (\ref{fff2}) for decays into 
a heavy and a light meson has been given at the two-loop 
order~\cite{BBNS00a}. Some of the arguments used there have 
straightforward extensions to all orders, but a technical 
all-order ``proof'' has not yet been accomplished. Nonetheless, 
the arguments used to prove infrared finiteness at two-loop order 
may be sufficiently convincing to make infrared finiteness at 
all orders plausible.

It has to be demonstrated that the hard-scattering kernels 
are infrared finite. To state this more precisely, we
write the factorization formula for a heavy-light final state 
schematically as
\begin{equation}
A \equiv  \langle \pi^- D^+|{\cal O}|\bar{B}_d\rangle 
= F_{B\to D}(0)\cdot T*\Phi_\pi,
\label{eq:factform}
\end{equation}
where the $*$ represents the convolution and ${\cal O}$ represents a 
four-quark operator. In order to extract $T$,
one computes $A$, $F_{B\to D}$ and $\Phi_\pi$ in perturbation theory and
uses (\ref{eq:factform}) to determine $T$. We therefore
rewrite (\ref{eq:factform}) in perturbation theory, 
\begin{eqnarray}
A^{(0)} + A^{(1)} +A^{(2)}+\,\cdots
&=&
\nonumber\\[0.1cm] 
&&\hspace{-3.5cm}
\left(F_{B\to D}^{(0)}+F_{B\to D}^{(1)}+F_{B\to D}^{(2)}+\cdots\right)
\nonumber\\[0.1cm]
&&\hspace{-3.5cm}\cdot\left(T^{(0)}+T^{(1)}+T^{(2)}+\cdots\right)
\nonumber\\[0.1cm] 
&&\hspace{-3.5cm}*\left(\Phi_\pi^{(0)}+\Phi_\pi^{(1)}+\Phi_\pi^{(2)}+\cdots\right),
\label{eq:ffpert}
\end{eqnarray}
where the superscripts in parentheses indicate the order of perturbation
theory, and then compare terms of the same order. Thus up to 
two-loop order
\begin{eqnarray}
\hspace*{-0.0cm}F_{B\to D}^{(0)}\cdot T^{(0)}*\Phi_\pi^{(0)}& = & 
A^{(0)},
\label{eq:t20}\\
\hspace*{-0.0cm}F_{B\to D}^{(0)}\cdot T^{(1)}*\Phi_\pi^{(0)}& = &
A^{(1)} 
\\
&&\hspace*{-1.2in}
-\,F_{B\to D}^{(1)}\cdot T^{(0)}*\Phi_\pi^{(0)}
-F_{B\to D}^{(0)}\cdot T^{(0)}*\Phi_\pi^{(1)},
\nonumber\label{eq:t21}\\
\hspace*{-0.0cm}F_{B\to D}^{(0)}\cdot T^{(2)}*\Phi_\pi^{(0)}& = &
A^{(2)}
\nonumber\\
&&\hspace*{-1.2in}
-\,F_{B\to D}^{(0)}\cdot T^{(1)}*\Phi_\pi^{(1)}
-F_{B\to D}^{(1)}\cdot T^{(1)}*\Phi_\pi^{(0)}
\nonumber\\ & & \hspace{-1.2in}
-F_{B\to D}^{(2)}\cdot T^{(0)}*\Phi_\pi^{(0)}
-F_{B\to D}^{(0)}\cdot T^{(0)}*\Phi_\pi^{(2)}
\nonumber\\ & & \hspace{-1.2in}
-F_{B\to D}^{(1)}\cdot T^{(0)}*\Phi_\pi^{(1)}.
\label{eq:t22}
\end{eqnarray}
By perturbative expansion of the $B\to D$ form factor, we mean the
perturbative expansion of the matrix element of $\bar{c}\Gamma b$,
evaluated between on-shell $b$- and $c$-quark states. By perturbative 
expansion of the pion light-cone distribution amplitude, we mean 
the perturbative expansion of the light-cone matrix element which 
defines the LCDA, but with the pion state replaced by an on-shell 
quark with momentum $u q$ and an on-shell antiquark with momentum 
$\bar{u} q$.

The zeroth order term in (\ref{eq:t20}) is trivial. The two terms 
that need to be subtracted from $A^{(1)}$ at first order exactly cancel 
the ``factorizable'' contributions to $A^{(1)}$. 
The first order term in (\ref{eq:t21}) therefore 
states that $T^{(1)}$ is given by the ``non-factorizable'' diagrams. 
(Here we use ``non-factorizable'' in the conventional sense, i.e. to 
denote the diagrams with gluon exchange between the $(BD)$ system and 
the pion.) If $T^{(1)}$ is to be infrared finite, the 
sum of these diagrams must be infrared finite, which is indeed the 
case as seen by the explicit one-loop calculation.

The second order term (\ref{eq:t22}) has a more complicated
structure. The last three terms on the right-hand side exactly cancel 
the ``factorizable'' corrections to the two-loop amplitude 
$A^{(2)}$. The remaining two terms that need to be subtracted from 
$A^{(2)}$ are non-trivial. The infrared divergences 
in the sum of ``non-factorizable'' contributions to $A^{(2)}$ must
then be shown to have 
precisely the right structure to match the infrared divergences 
in $F_{B\to D}^{(1)}$ and $\Phi_\pi^{(1)}$, such that
\begin{eqnarray}
\label{cc1}
&&\hspace*{-0.6cm}
A^{(2)}_{\rm non-fact.}
-F_{B\to D}^{(0)}\cdot T^{(1)}*\Phi_\pi^{(1)}
\nonumber\\
&& \hspace*{-0.6cm}-\,F_{B\to D}^{(1)}\cdot T^{(1)}*\Phi_\pi^{(0)} 
= 
\mbox{ infrared finite.}
\end{eqnarray}
Eq.~(\ref{cc1}) is verified by first identifying the regions
of phase space which can give rise to infrared singularities. In
general these arise when massless lines become soft or 
collinear with the direction of $q$, the momentum of the pion. This 
requires that one or both of the loop momenta in a two-loop diagram 
become soft or collinear. Rather than computing the 62 
``non-factorizable'' two-loop diagrams (excluding self-energy
insertions and field renormalization),
the Feynman integrands corresponding to 
these diagrams in those momentum configurations that can 
give rise to singularities are then analyzed in all possible
combinations: one momentum soft or collinear, the other hard; 
both momenta soft 
or collinear; one momentum soft, the other collinear. 

The analysis of Ref.~\cite{BBNS00a} shows that infrared 
divergences cancel in the soft-soft, collinear-col\-linear and 
soft-collinear region as required for the validity of (\ref{cc1}). 
The non-cancelling divergence in the hard-soft region factorizes 
into the form
\begin{equation}
A^{(2)}_{\rm hard-soft} = f_{B\to D}\cdot 
T^{(1)}*\Phi_\pi^{(0)},
\end{equation}
where $f_{B\to D}$ is precisely the soft contribution to the 
$B\to D$ form factor at the one-loop order; 
this cancels the second of the two subtraction 
terms in (\ref{cc1}). The infrared divergent hard-collinear 
contributions sum up to the expression 
\begin{eqnarray}\label{stvphi}
A^{(2)}_{\rm hard-collinear}&=&F_{B\to D}^{(0)}\cdot
C_F\frac{\alpha_s}{\pi}\ln\frac{\mu_{UV}}{\mu_{IR}}
\nonumber\\
&&\hspace*{-2.5cm}
\times\,\int^1_0\, dw\, du\, T^{(1)}(w)\, V(w,u)\, \Phi^{(0)}_\pi(u),
\end{eqnarray}
with $V(w,u)$ the ERBL evolution kernel~\cite{LB80,EfRa80}. 
The infrared singular contribution to the (perturbative, one-loop) 
pion distribution amplitude is determined by  
\begin{equation}
\label{eq:erbl1}
\Phi^{(1)}_\pi(w) = 
C_F\,\frac{\alpha_s}{\pi}\,\ln\frac{\mu_{UV}}{\mu_{IR}}\,
\int_0^1 du\,V(w,u)\,\Phi^{(0)}_\pi(u),
\end{equation}
and by combining the previous two equations we find that 
$A^{(2)}_{\rm hard-collinear}$ is precisely equal to the 
remaining subtraction term in (\ref{cc1}). It follows from 
(\ref{eq:t22}) that 
$T^{(2)}$ is free of infrared singularities. 

\subsection{Rescattering and Parton-Hadron Duality}
\label{rescatt}

Final-state interactions are usually discussed in terms of intermediate 
hadronic states. This is suggested by the unitarity relation (taking 
$B\to \pi\pi$ for definiteness) 
\begin{equation}
\label{unitarity}
\mbox{Im}\,{\cal A}_{B\to \pi\pi} \sim \sum_n 
{\cal A}_{B\to n}{\cal A}_{n\to \pi\pi}^*,
\end{equation}
where $n$ runs over all hadronic intermediate states. In many 
discussions of final state rescattering the sum on the 
right hand side of this equation is truncated by keeping only 
elastic rescattering. It is clear that this approximation is 
incompatible with the heavy-quark limit, in which the opposite limit 
of an arbitrarily large number of intermediate states should 
be considered. Decays of $B$ mesons lie somewhere in between 
these limiting cases, but only the heavy-quark limit provides 
a controlled approximation to the problem. In my opinion, in view 
of the factorization results, proponents of the elastic scattering 
limit and methods inspired by Regge physics now need to justify better 
their motivation for choosing this particular ansatz.

The heavy-quark limit, and the dominance of hard rescattering in 
this limit, suggest that the sum in (\ref{unitarity}) is interpreted 
as extending over intermediate states of partons. In this picture 
the sum over all hadronic intermediate states is approximately 
equal to the contribution of a quark-anti-quark intermediate state 
of small transverse extension. The approximation could be further 
improved by including $q\bar{q}g$ intermediate states etc. This 
is similar to the QCD description of $e^+ e^-\to\,$hadrons at 
large energy; the total cross section of this reaction is well
represented by the production cross section of a $q\bar{q}$ pair, 
even though the production of every particular final state cannot 
be computed with perturbative (or any known) methods. There is a limit
to the accuracy of such kinds of descriptions, which is discussed 
under the name of ``parton-hadron'' duality. Quantifying this 
accuracy is a formidable, unsolved problem.  The same assumption
forms the basis for the application of the operator product expansion to 
{\em inclusive\/} non-leptonic heavy-quark decays and there have been 
some quantitative studies in this context, though in the  
two-dimensional 't~Hooft model~\cite{GL97,Bigi99,Lebed:2000gm}.  
Parton-hadron duality is also an implicit assumption in applying 
perturbative QCD techniques to jet observables 
and hadron-hadron collisions at large momentum transfer. It is 
probably safe to conclude that the accumulated experience 
suggests that violations of parton-duality are subdominant 
effects in the heavy-quark limit, and this is all we need to 
justify our theoretical framework.

\subsection{Higher Fock States and Non-factori\-zable Contributions}
\label{higherfock}

The factorization formula needs only the leading-twist LCDAs of 
the mesons. Higher Fock components of the mesons appear in higher 
orders of the collinear expansion. The collinear expansion is 
justified as long as the additional partons carry a finite fraction 
of the meson's momentum in the heavy-quark limit. 
Under this assumption, it is easy to see that 
adding additional partons to the Fock state increases the number 
of off-shell propagators in a given diagram. This implies power
suppression in the heavy-quark expansion.

Soft contributions are also power-suppressed, but it seems difficult 
to classify them systematically. Again the decay 
$\bar{B}_d\to D^+ \pi^-$ is the simplest case, and I briefly consider 
the situation, where the ``non-factorizable'' gluon, 
i.e.\ the gluon exchanged between the pion and the $(\bar{B}D)$ 
system, is soft. In this case, the ``$q\bar qg$ Fock state'' cannot be
described by a light-cone wave function, but such a
contribution still receives a power suppression in the 
heavy-quark limit. The important point is that soft gluons couple 
very weakly to the $q\bar{q}$ pair. The coupling can be evaluated 
(the $q\bar{q}$ pair being very energetic), and the result is  
\begin{eqnarray}
\label{o8snf}
&&\hspace*{-0.3cm}\langle D^+\pi^-|(\bar{c}T^Ab)_{V-A}(\bar{d}T^Au)_{V-A}|
\bar B_d\rangle_{\rm nf} = 
\nonumber \\[0.1cm]
&&\hspace*{-0.1cm}
-\int^{1/\Lambda_{\rm QCD}}_0 \hspace*{-0.8cm}
ds\,\langle D^+|\bar c\gamma^\mu(1-\gamma_5)
g_s\tilde G_{\mu\nu}(-s n) n^\nu b|\bar B_d\rangle\,
\nonumber\\
&&\times\,
\int^1_0 du\,\frac{f_\pi\Phi_\pi(u)}{8N_c u\bar u},
\end{eqnarray}
where $q=E n$ is the momentum of the pion. This depends on a 
{\em non-local} higher-dimension $B\to D$ form factor, but comparing 
this with the leading order, conventional factorization 
expression,
\begin{eqnarray}\label{o1lf}
&&\hspace*{-0.3cm}
\langle D^+\pi^-|(\bar{c}b)_{V-A}(\bar{d}u)_{V-A}|\bar B_d\rangle_{\rm LO}=
\\
&&\langle D^+|\bar c\gamma^\mu(1-\gamma_5)b|\bar B_d\rangle\,
i f_\pi E n_\mu \!\int^1_0 \!du\,\Phi_\pi(u),
\nonumber\end{eqnarray}
we conclude on dimensional arguments that the soft non-factorizable 
correction is suppressed by one power of $\Lambda_{\rm QCD}/m_b$. 
(Note that similar considerations for the $J/\psi K$ final 
state\,\cite{Khod98} lead to {\em local} $B\to K$ form factors, because 
the non-locality is then cut off at a distance $1/m_c$.)

Note that power corrections come without a factor of $\alpha_s$ and 
we may expect them to be as large as the computable perturbative 
corrections in general. An important point, however, is that there
exists a systematic framework that allows us to classify {\em both} 
effects as corrections.

\subsection{``Chirally enhanced'' Corrections}
\label{chiral}

There are two reasons why the hard spectator interaction in
(\ref{fff}) is particularly sensitive to power-suppressed
corrections. The first reason is that the short-distance scale is not 
$m_b$ (as is the case for the form factor term in (\ref{fff})), 
but $(m_b\Lambda_{\rm QCD})^{1/2}$. To see this, note the 
gluon virtuality is
\begin{equation}
k_g^2=-\bar{v}\xi M_B^2+\,\mbox{terms of order}\,\Lambda_{\rm QCD}^2,
\end{equation}
which on average is around $1\,\mbox{GeV}^2$. To arrive at 
(\ref{hard}) I have neglected the terms of order 
$\Lambda_{\rm QCD}^2$ and this might not be a particularly 
good approximation. However, there is no (known) systematic way 
of treating such terms, which amongst other things are sensitive 
to the off-shellness of the spectator quark in the $B$ meson 
(and hence to higher Fock components of the $B$ meson), and so we 
must neglect these terms together with many other power 
corrections.

There is an enhancement of power-suppressed effects for decays into
light pseudoscalar mesons connected with the curious numerical fact that 
\begin{equation}
\label{mupi}
2\mu_\pi\equiv 
\frac{2 m_\pi^2}{m_u+m_d} = -\frac{4 \langle\bar{q}q\rangle}{f_\pi^2} 
\approx 3\,\mbox{GeV}
\end{equation}
is much larger than its naive scaling estimate $\Lambda_{\rm QCD}$. 
(Here $\langle \bar{q}q\rangle = \langle 0|\bar{u} u|0\rangle= 
\langle 0|\bar{d} d|0\rangle$ is the 
quark condensate.) These ``chirally enhanced'' corrections have 
originally been discussed in connection with $V+A$ penguin 
operators in the weak effective hamiltonian, but they affect the 
hard spectator interaction more severely. 

Consider first the contribution of the operator 
${\cal O}_6=(\bar{d}_i b_j)_{V-A} (\bar{u}_j u_i)_{V+A}$ to the 
$\bar{B}_d\to\pi^+\pi^-$ decay amplitude. The parameter 
$\mu_\pi$ arises already in the naively factorized matrix element: 
\begin{eqnarray}
\label{lotw3}
\hspace*{-1cm}\langle \pi^+\pi^-|(\bar{d}_i b_j)_{V-A} (\bar{u}_j u_i)_{V+A}|
\bar{B}_d\rangle &=&
\nonumber\\[0.1cm]
\hspace*{0cm}i M_B^2 F_+^{B\to \pi}(0) f_\pi \times \frac{2\mu_\pi}{m_b}.
\end{eqnarray}
This is formally a $\Lambda_{\rm QCD}/m_b$ power correction 
but numerically large due to (\ref{mupi}).
We would not have to worry about such terms if they could all be 
identified and the factorization formula (\ref{fff}) applied 
to them, since in this case higher-order perturbative 
corrections would not contain non-factorizing
infrared logarithms. However, this is not the case.

After including radiative corrections, the matrix element on the 
left-hand side of (\ref{lotw3}) is expressed as a non-trivial 
convolution with the pion light-cone 
distribution amplitude. The terms involving $\mu_\pi$ can be 
related to two-particle twist-3 (rather than leading twist-2) 
distribution amplitudes, conventionally called $\Phi_p(u)$ and 
$\Phi_\sigma(u)$. The distribution amplitude $\Phi_p(u)$ does not 
vanish at the endpoint. As a consequence the hard spectator interaction 
contains an endpoint divergence. In other words, 
the ``correction'' relative to (\ref{lotw3}) is of the form 
$\alpha_s\times\,$logarithmic divergence, which we interpret as 
being of the same order as (\ref{lotw3}). It turns out, however, 
that the $\alpha_s$ correction to the $V+A$ operator (giving rise 
to the parameter $a_6$ in conventional notation) is free of this 
potential problem. 

As a consequence the most important effect of the chirally enhanced 
twist-3 LCDAs (with exception of the leading order result for $a_6$) 
is on the matrix elements that contribute to the 
coefficients $a_1$ to $a_5$. An example of this is shown in 
(\ref{hard}). Substituting the asymptotic LCDAs, $H$ is rewritten 
as 
\begin{equation}
\label{hard2}
H = \frac{12\pi^2\,f_B f_\pi}{M_B \lambda_B F^{B\to \pi}_+(0)}
\left[1 + \frac{2\mu_\pi}{3 m_b}\int_0^1 \frac{d v}{v}\right], 
\end{equation}
which exhibits the problem of dealing with the endpoint-divergent 
integral. 
I should emphasize that this divergence is not in contradiction 
with the factorization formula (\ref{fff}), in fact it is expected 
at the level of power-suppressed effects. But from the
phenomenological point of view it is somewhat disappointing that 
these effects are sizeable and can introduce a substantial uncertainty
into the hard spectator interaction. The complete set of chirally enhanced 
terms has been estimated up to now only in Ref.~\cite{BBNS00b}, 
where it was assumed that the divergent integral can be replaced 
by a universal constant. The variation of this constant constitutes 
the largest theoretical uncertainty, but it is also shown that 
the predictivity of the approach is not lost. 
(Some chirally enhanced terms have been included in 
Refs.~\cite{MSYY00,DYZ00}, but the correction (\ref{hard}), (\ref{hard2}) 
to $a_1$ to $a_5$, which contains the endpoint divergence, 
has been omitted~\cite{MSYY00} or computed incorrectly~\cite{DYZ00} 
in these papers.)
As in a related situation for the pion form factor 
\cite{GT82} one might argue that the endpoint divergence is suppressed 
by a Sudakov form factor. However, it is likely that when $m_b$ 
is not large enough to suppress these chirally enhanced terms, then it is 
also not large enough to make Sudakov suppression effective especially
since the short-distance scale is not large enough to build up a 
strong logarithmic evolution.

\subsection{Annihilation Topologies}
\label{annihilation}

\begin{figure}[t]
   \vspace{-2.6cm}
   \epsfysize=13.5cm
   \epsfxsize=9cm
   \centerline{\epsffile{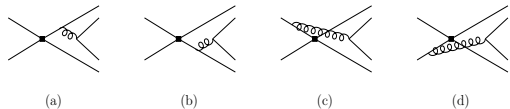}}
   \vspace*{-9.2cm}
\caption[dummy]{\label{fig9}\small Annihilation diagrams.}
\end{figure}

My final concern in this section are the annihilation topologies 
(Fig.~\ref{fig9}). The hard part of these diagrams would 
amount to another contribution to the second hard-scattering kernel, 
$T^{II}_i(\xi,u,v)$, in (\ref{fff}). The soft part, if unsuppressed, 
would violate factorization. However, a straightforward power-counting
analysis shows that all annihilation topologies are $1/m_b$ 
corrections to the decay amplitudes in the heavy-quark 
limit~\cite{BBNS00a}. This statement also applies to diagram d, 
in which case the largest term comes from an endpoint 
contribution.

As for the hard spectator interaction at leading power in the 
heavy-quark expansion, there exist ``chirally enhanced'' contributions
from the annihilation topologies related to the corresponding twist-3 
light meson LCDAs. It has recently been 
noted~\cite{Keum:2000wi,Lu:2000em} in the context of the
hard-scattering approach that these could be
non-neg\-li\-gible. To illustrate this effect, I consider the annihilation
correction to the coefficient $a_6$ in the effective transition 
operator defined in Ref.~\cite{BBNS99}. Note that $a_6$ is the
coefficient of a 
power correction (though chirally enhanced), and I am considering 
now a power correction to $a_6$. For two identical final state 
mesons, say two pions, only the diagrams 
a and b contribute to $a_6$. To simplify the
result, I assume the LCDAs to be the asymptotic ones and obtain 
for the sum of leading order and annihilation contribution: 
\begin{eqnarray}
a_6 &\simeq& \left(C_6+\frac{C_5}{N_c}\right)\Bigg[1+\frac{\alpha_s}{9\pi}\,
\frac{96\pi^2 f_B f_\pi}{M_B^2 F^{B\to\pi}_+(0)} 
\nonumber\\
&&\hspace*{0cm}
\times\,\left(X_l^2-\frac{X_l}{2}\right)\bigg],
\end{eqnarray}
where $X_l$ is the divergent integral $\int_0^1 dv/v$. 
Al\-though power-suppressed, the correction is enlarged by a numerical 
factor and a logarithmic 
endpoint divergence from each of the two final mesons. 
We can exhibit this more transparently by comparing the annihilation 
correction to the generic leading-power hard spectator correction 
(\ref{hard}), (\ref{hard2}). This gives the ratio 
\begin{equation}
\frac{H_{\rm ann}}{H} \simeq \frac{\lambda_B}{M_B} 
\times 8 \left(X_l^2-\frac{X_l}{2}\right),
\end{equation}
suggesting that in this particular case the annihilation topologies 
are more important than the generic hard spectator interaction. A complete 
analysis of annihilation contributions to light-light final states 
(extending the analysis~\cite{BBNS00a} for the $D \pi$ case) is 
currently in progress.

\section{Conclusion} 

The QCD factorization approach described in this talk constitutes 
a powerful and systematic approach to non-leptonic decay amplitudes, 
based on familiar methods of perturbative QCD, and the assumption 
that the $b$ quark mass is large. It does not render trivial the 
problem of accurately computing these amplitudes, but it appears  
to me that the outstanding issues now are more of numerical   
than of conceptual character: the best way of dealing with chirally
enhanced power corrections; the role of annihilation; the size of 
power corrections in general and their impact on strong interaction 
phases; the role of hard scattering in heavy-to-light form 
factors (and, related to this, the importance of Sudakov form 
factors) $\ldots$. It is probable that experimental data will be needed 
to shed light on some of these issues.

\section*{Acknowledgments}
I would like to thank G.~Buchalla, M.~Neubert and C.T.~Sachrajda for 
their collaboration on the topics of this talk. I would also like 
to thank the organizers of the conference for their generous 
support.

\end{document}